\documentstyle[twoside,fleqn,amssymb,espcrc2]{article}

\newcommand{\AmS}{{\protect\the\textfont2
  A\kern-.1667em\lower.5ex\hbox{M}\kern-.125emS}}
\newcommand{\be}{\begin{equation}}
\newcommand{\ee}{\end{equation}}
\newcommand{\bea}{\begin{eqnarray}}
\newcommand{\eea}{\end{eqnarray}}
\def\pmas{\partial_+}
\def\pmen{\partial_-}
\def\Nu{{\cal{V}}}
\def\a{\alpha}
\def\b{\beta}
\def\g{\gamma}
\def\e{\epsilon}

\title{Integrable models, degenerate horizons and AdS$_2$ black holes}
\author{J. Cruz,\address{Departamento de F\'{\i}sica Te\'orica and
        IFIC, Centro Mixto Universidad de Valencia-CSIC,\\
        Facultad de F\'{\i}sica, Universidad de Valencia,
        Burjassot-46100, Valencia, Spain.}
        A. Fabbri,\address{Dipartimento di Fisica dell'Universit\`a di
    Bologna and INFN sezione di Bologna, Via Irnerio 46, 40126 Bologna, Italy.}
    D. J. Navarro$^{\rm a}$, J. Navarro-Salas$^{\rm a}$ and
    P. Navarro$^{\rm a}$}
\begin{document}

\begin{abstract}
The near extremal Reissner-Nordstr\"om black holes in arbitrary dimensions can
be modeled by the Jackiw-Teitelboim (JT) theory. The asymptotic Virasoro
symmetry of the corresponding JT model exactly reproduces, via Cardy's formula,
the deviation of the Bekenstein-Hawking entropy of the Reissner-Nordstr\"om
black holes from extremality. We also comment how can we extend this approach
to investigate the evaporation process.
\end{abstract}

\maketitle

The understanding of the thermodynamical properties and the dynamical evolution
of black holes is a crucial aspect in the formulation of a consistent theory of
quantum gravity. Recent improvements of string theory have been able to provide
a microscopical structure which accounts for the Bekenstein-Hawking entropy and
gives a mechanism for the evaporation process of a black hole (see the review
\cite{ma}). More recently, the Bekenstein-Hawking entropy of three-dimensional
BTZ black holes has been derived using only symmetry properties of
three-dimensional gravity \cite{st} (in this approach there are some technical
subtleties \cite{c1}). This suggests that the entropy formula may be insensitive
to the details of the quantum theory. Further results in this direction can be
found in \cite{c2,so}.\\

An extremal 5d Reissner-Nordstr\"om black hole was the first case
in which it was possible to identify the microscopic degrees of freedom
responsible for the Bekenstein-Hawking entropy \cite{sv}. The extremality
condition was crucial since it corresponds, although this link is not general,
to a BPS state. The deviation of the near-extremal entropy \cite{cm} is also
related to the BPS bound. In this work we want to point out that, since the
extremality condition is also related to the existence of degenerate horizons
and hence to two-dimensional Anti-de Sitter geometry, a realization of the
AdS$_2$/CFT$_1$ correspondence \cite{nn,nnn} can be used to explain the
near-extremal Bekenstein-Hawking entropy of Reissner-Nordstr\"om black holes in
arbitrary dimensions. Moreover we argue that the two-dimensional effective
theory which codifies the dynamics near extremality could also be used to provide a
picture of the evaporation process.\\

Let us start our analysis with the Einstein-Maxwell action in $(n+2)$
dimensions
\bea
\label{naction}
I^{(n+2)} &=& \frac{1}{16\pi l^n} \int d^{n+2}x \sqrt{-g^{(n+2)}} \left(
R^{(n+2)} \right. \> \nonumber \\
&+& \left. (F^{(n+2)})^2 \right) \, ,
\eea
where $l^n$ is Newton's constant $G^{(n+2)}$. Imposing spherical symmetry on
the gauge field and decomposing the metric in the form
\bea
ds^2_{(n+2)} &=& \frac{8(n-1)}{n^2} \left( \frac{8(n-1)}{n} \phi
\right)^{\frac{n}{n-1}} ds^2_{(2)} \nonumber \\
&+& l^2 \left( \frac{8(n-1)}{n} \phi \right)^{\frac{2}{n}}
d\Omega^2_{(n)} \, ,
\eea
the above action reduces to
\be
\label{2Daction}
I^{(n+2)} = \frac{1}{2G} \int d^2x \sqrt{-g} (R\phi + l^{-2} V(\phi)) \, ,
\ee
where
\be
G = \frac{n\pi}{(n-1) \Nu^{(n)}} \, ,
\ee
being $\Nu^{(n)}$ the area of the unit S$^n$ sphere and
\bea
V(\phi) &=& (n-1) \left( \frac{8(n-1)}{n} \phi \right)^{\frac{-1}{n}} \\
&-&  \frac{2(n-1)^2l^2q^2}{n} \left( \frac{8(n-1)}{n} \phi
\right)^{\frac{1-2n}{n}} \, . \nonumber
\eea
In terms of the two-dimensional metric the Reissner-Nordstr\"om black hole
solutions take the form
\bea
\label{2metric}
ds_{(2)}^2 &=& -U(\phi) dt^2 +U(\phi)^{-1} dx^2 \, , \nonumber \\
U(\phi) &=& J(\phi)-2Glm \, , \\
\phi &=& \frac{x}{l} \, , \nonumber
\eea
where $J(\phi)=\int_0^{\phi} d\tilde{\phi} V(\tilde{\phi})$. The horizons of
the black hole correspond to
\be
J(\phi) = 2Glm \, ,
\ee
and the degenerate horizons emerge when
\be
J^{\prime}(\phi_0) = 0 \, .
\ee
The idea now is to perform a perturbation around the degenerate radius of
coincident horizons \cite{cfnn}
\bea
m &=& m_0(1+k\a^2) \, , \\
t &=& \frac{\tilde{t}}{\a} \, , \\
x &=& x_0 + \a \tilde{x} \, , \\
\phi &=& \phi_0 + \a \tilde{\phi} \, ,
\eea
where $J(\phi_0)=2Glm_0$. We then have
\bea
\label{nearmetric}
ds^2 &=& -\tilde{U}(r) d\tilde{t}^2 +
\frac{d\tilde{x}^2}{\tilde{U}(r)} + r_0^2 d\Omega^2 + \cal{O}(\a) \, ,
\nonumber \\
\tilde{U}(r) &=& -\frac{\tilde{R}_0}{2} \tilde{x}^2 - km_0l \, ,
\eea
where
\be
\tilde{R}_0 = -\frac{J^{\prime\prime}(\phi_0)}{l^2} \, .
\ee
The two-dimensional effective action can also be expanded
\bea
\label{effective}
I^{(n+2)} &=& \frac{\a}{2G} \int d^2x \sqrt{-g} (R\tilde{\phi} \nonumber \\
&+& l^{-2} V^{\prime}(\phi_0) \tilde{\phi}) + {\cal{O}}(\a^2) \, ,
\eea
and the leading order term is just the Jackiw-Teitelboim model. This theory
possesses an asymptotic set of symmetries \cite{cami,nnn} which preserve the
asymptotic form of the metric
\bea
g_{\tilde{t}\tilde{t}} &=& \phantom{-} \frac{\tilde{R}_0}{2} \tilde{x}^2 +
\g_{\tilde{t}\tilde{t}} + \ldots \> , \\
g_{\tilde{t}\tilde{x}} &=& \phantom{-} \frac{\g_{\tilde{t}\tilde{x}}}
{\tilde{x}^3} + \ldots \> , \\
g_{\tilde{x}\tilde{x}} &=& -\frac{2}{\tilde{R}_0} \frac{1}{\tilde{x}^2} +
\frac{\g_{\tilde{x}\tilde{x}}}{\tilde{x}^4} + \ldots \> .
\eea
The infinitesimal diffeomorphisms $\zeta^a(\tilde{x}, \tilde{t})$ preserving
the above boundary conditions are
\bea
\zeta^{\tilde{t}} &=& \epsilon(\tilde{t}) - \frac{2}{\tilde{R}_0^2\tilde{x}^2}
\epsilon(\tilde{t}) + {\cal{O}}\left( \frac{1}{\tilde{x}^4} \right) \, , \\
\zeta^{\tilde{x}} &=& -\tilde{x} \epsilon^{\prime}(\tilde{t}) +
{\cal{O}}\left(\frac{1}{\tilde{x}} \right) \, ,
\eea
where $\epsilon(\tilde{t})$ is an arbitrary function. The remaining terms in
the above expansion represent pure gauge transformations. The following
expression
\be
\Theta_{\tilde{t}\tilde{t}}=\kappa \left( \gamma_{\tilde{t}\tilde{t}}-
\frac{1}{2} \left( \frac{\tilde{R}_0}{2} \right)^2 \gamma_{\tilde{x}\tilde{x}}
\right) \, ,
\ee
where $\kappa$ is a model-dependent constant, is the unique ``gauge invariant''
quantity and it transforms according to the rule
\be
\label{rule}
\delta_{\epsilon}\Theta_{\tilde{t}\tilde{t}} = \epsilon(\tilde{t})
\Theta_{\tilde{t}\tilde{t}}^{\prime} + 2\Theta_{\tilde{t}\tilde{t}}
\epsilon^{\prime}(\tilde{t}) - \frac{2\kappa}{\tilde{R}_0} \epsilon^{\prime
\prime\prime}(\tilde{t}) \, .
\ee
Since Anti-de Sitter space has a natural periodicity in $\tilde{t}$ we shall
assume that it varies in the interval $0 \leq \tilde{t} \leq 2 \pi \b$. Defining the
Fourier components of $\Theta_{\tilde{t}\tilde{t}}$ as
\be
\label{mode}
L_n^R = \frac{1}{2\pi \b} \int_0^{2\pi \b} d\tilde{t}
\Theta_{\tilde{t}\tilde{t}} \b e^{in\frac{\tilde{t}}{\b}} \, ,
\ee
the Poisson algebra can be computed from
\be
\{ L_n, L_m \} = \delta_{\e_m} L_n \, ,
\ee
where $\e_m=\b e^{im\frac{\tilde{t}}{\b}}$, and the result is a Virasoro
algebra with central charge
\be
c =- \frac{24}{\tilde{R}_0 \b} \kappa \, .
\ee
The coefficient $\kappa$ is determined by the effective Lagrangian near
extremality and can be computed using Hamiltonian methods. We find that
$\kappa=\frac{\a}{lG}$, so
\be
c = -\frac{24\a}{lG\tilde{R}_0\b} \, .
\ee
One can also work out the value of $L_0$
\be
L_0 = m_0 \kappa \a \b \, .
\ee

We must remark here that in the Fourier decomposition the parameter
$\tilde{t}$ plays the role of a null coordinate of a 2d CFT. This is a
consequence of the general fact \cite{st2} that the conformal group in the
AdS$_2$/CFT$_1$ correspondence can be regarded as one chiral component of the
two-dimensional conformal group. With these values the product $cL_0$ which
appears in the Cardy formula is insensitive to the arbitrary parameters
introduced so far and only depends on physical quantities. We have
\be
2\pi \sqrt{\frac{cL_0}{6}} = 2\pi \sqrt{\frac{4\Delta m}{-\tilde{R}_0lG}} \, ,
\ee
and taking into account that
\be
\tilde{R}_0 = \frac{16(n-1)^3}{-l^2n^2} \left(
\frac{n}{2(n-1)l^2q^2} \right)^{\frac{1+n}{2(n-1)}} \, ,
\ee
the above expression exactly coincides with the deviation $\Delta S$ of the
Bekenstein-Hawking entropy
\be
S^{BH} = \frac{\Nu^{(n)} r^n}{4l^n} \, ,
\ee
from the extremal case. Therefore, two-dimensional gravity theories codify
adequately the effective dynamics in such a way that they allow to provide a
microscopic derivation of the near-extremal black hole entropy. So it is also
natural to investigate whether the Jackiw-Teitelboim model can provide a
mechanism for the evaporation process of a Reissner-Nordstr\"om black hole near
extremality.\\

We shall now mention briefly how one can investigate the evaporation of
Reissner-Nordstr\"om black holes using the approximation (\ref{effective}). The
Jackiw-Teitelboim theory is an integrable 2d dilaton gravity model \cite{f}
with general solution given by the expression \cite{f,cinn} (here we introduce
$\lambda^2 = \frac{V^{\prime}(\phi_0)}{4l^2}$)
\bea
ds^2 &=& -\frac{\pmas A_+ \pmen A_-}{(1+\frac{\lambda^2}{2} A_+ A_-)^2} \, , \\
\phi &=& -\frac{1}{2} \left( \frac{\pmas \tilde{a}_+}{\pmas A_+} +
\frac{\pmen \tilde{a}_-}{\pmen A_-} \right) \nonumber \\
&+& \frac{\lambda^2}{2} \frac{A_+\tilde{a}_- + A_- \tilde{a}_+}{1+\frac{\lambda^2}{2}
A_+ A_-} \, ,
\eea
where the chiral functions $A_{\pm}$, $a_{\pm}$ verify the constraint equations
\bea
T^f_{\pm\pm} &=& \pm \partial_{\pm}^2 \left( \frac{\partial_{\pm} a_{\pm}}
{\partial_{\pm} A_{\pm}} \right) \nonumber \\
&\mp& \frac{\partial_{\pm}^2
A_{\pm}}{\partial_{\pm} A_{\pm}} \; \partial{\pm} \left( \frac{\partial_{\pm}
a_{\pm}}{\partial_{\pm} A_{\pm}} \right)  \, .
\eea
Working in the conformal gauge we can match along the null line $x^+=x_0^+$ the
solution $\phi=0$ ($a_{\pm}=0$),
representing an extremal black hole,
with the
solution $\phi \neq 0$
\be
\phi = C \frac{\frac{\lambda^2}{2}(x^+ \Delta^- + x^- \Delta^+) - 1 +
\frac{\lambda^2}{2} x^+x^-}{1 + \frac{\lambda^2}{2} x^+x^-} \, ,
\ee
where $\Delta^+=-x_0^+$, $\Delta^-={\frac{2}{\lambda^2x_0^+}}$ and
\be
ds^2 = \frac{-dx^+dx^-}{(1 + \frac{\lambda^2}{2} x^+x^-)^2} \, ,
\ee
everywhere.
The above dynamics is originated by the shock wave
\be
T^f_{++} = \frac{C}{x_0^+} \delta (x^+-x_0^+) \,
\ee
and therefore  it seems natural to associate
the solution for $x^+>x_0^+$ to a near extremal hole.\\

The solvability of the classical model when matter is coupled in a conformal
way can also be extended to the one-loop theory. The effective action is given
by
\bea
S &=& \frac{1}{2\pi} \int d^2x \sqrt{-g} \left(R \tilde{\phi} + 4 \lambda^2
\tilde{\phi} \right. \nonumber \\
&-& \left. \sum_{i=1}^N \frac{1}{2} |\nabla f_i|^2\right) - \frac{N}{96\pi}
\int \sqrt{-g} R \square^{-1} R \nonumber \\
&+& \frac{N}{12\pi} \int d^2x \sqrt{-g} \lambda^2 \, ,
\eea
where the $N$ fields $f_i$ model the matter degrees of freedom.\\

The unconstrained equations of motion remain as the classical
ones, but the constrained equations get modified according to \bea
& T^f_{\pm\pm} - \frac{N}{12} t_{\pm} = & \nonumber \\ &\pm
\frac{1}{2} \partial_{\pm}^2 \left( \frac{\partial_{\pm} a_{\pm}}
{\partial_{\pm} A_{\pm}} \right) \mp  \frac{1}{2}
\frac{\partial_{\pm}^2 A_{\pm}}{\partial_{\pm} A_{\pm}} \;
\partial{\pm} \left( \frac{\partial_{\pm} a_{\pm}}{\partial_{\pm}
A_{\pm}} \right) & \nonumber \\ & + \frac{N}{12} \left[
\frac{1}{4} \left( \frac{\partial^2_{\pm} A_{\pm}} {\partial_{\pm}
A_{\pm}} \right)^2 - \frac{1}{2} \; \partial_{\pm} \left(
\frac{\partial^2_{\pm} A_{\pm}}{\partial_{\pm} A_{\pm}} \right)
\right] & \, , \eea where the functions $t_{\pm}(x^{\pm})$ are
related with the boundary conditions of the theory. The crucial
point now is to choose the proper physical boundary functions
$t_{\pm}$. This question will be investigated in a further
work, but we want to stress that, if the JT model has been able to
explain the deviation of the near extremal entropy one can also
expect that a rigorous study would offer an approximate but useful
picture of the evaporation process of near-extremal
Reissner-Nordstr\"om black holes.

\section*{Acknowledgements}
This research has been partially supported by the CICYT and DGICYT, Spain.
J. Cruz acknowledges the Generalitat Valenciana for a FPI fellowship.
D. J. Navarro thanks the Ministerio de Educaci\'on y Cultura for a FPI
fellowship. P. Navarro acknowledges the Ministerio de Educaci\'on y Cultura
for a FPU fellowship.

\end{document}